\pdfoutput=1
\documentclass{PoS}

\usepackage{wrapfig}
\usepackage{multirow}
\usepackage{graphicx}
\usepackage{units}
\usepackage{nicefrac}
\usepackage{amsmath}

\usepackage[font=footnotesize,labelfont=bf]{caption}
\usepackage[font=footnotesize,labelfont=bf]{subcaption}

\DeclareCaptionSubType*[alph]{table}
\graphicspath{{graphics/}}

\title{
\vspace{-2.5cm}{\scriptsize \rm \hfill HU-EP-14/53, SFB/CPP-14-91, DESY 14-218}\\
\vspace{2cm}
Progress in Simulations with Twisted Mass Fermions at the Physical Point}

\ShortTitle{mtmLQCD at the Physical Point}

\newcounter{institutecounter}
\edef\cyp{\alph{institutecounter}}
\stepcounter{institutecounter}
\edef\ucy{\alph{institutecounter}}
\stepcounter{institutecounter}
\edef\rmii{\alph{institutecounter}}
\stepcounter{institutecounter}
\edef\fer{\alph{institutecounter}}
\stepcounter{institutecounter}
\edef\nic{\alph{institutecounter}}
\stepcounter{institutecounter}
\edef\hub{\alph{institutecounter}}
\stepcounter{institutecounter}
\edef\gre{\alph{institutecounter}}
\stepcounter{institutecounter}
\edef\bon{\alph{institutecounter}}
\stepcounter{institutecounter}
\edef\ber{\alph{institutecounter}}

\author{ A.~Abdel-Rehim$^{(\cyp)}$,
      C.~Alexandrou$^{(\cyp,\ucy)}$,
      P.~Dimopoulos$^{(\rmii,\fer)}$,
      R.~Frezzotti$^{(\rmii)}$,
      K.~Jansen$^{(\nic)}$,
      C.~Kallidonis$^{(\cyp)}$,
      \speaker{B.~Kostrzewa}$^{(\hub,\nic)}$,
      M.~Mangin-Brinet$^{(\gre)}$,
      G.C.~Rossi$^{(\rmii,\fer)}$,
      C.~Urbach$^{(\bon)}$,
      U.~Wenger$^{(\ber)}$ \\

      $^{(\cyp)}$ CaSToRC, The Cyprus Institute, 2121 Aglantzia, Nicosia, Cyprus

      $^{(\ucy)}$ Department of Physics, University of Cyprus, P.O. Box 20537, 1678 Nicosia, Cyprus

      $^{(\rmii)}$ Dip. di Fisica, Universit{\`a} di Roma Tor Vergata and INFN, I-00133 Roma, Italy

      $^{(\fer)}$ Centro Fermi, Piazza del Viminale 1, I-00184 Rome, Italy

      $^{(\nic)}$ NIC, DESY, Zeuthen, Platanenallee 6, D-15738 Zeuthen, Germany

      $^{(\hub)}$ Institut f\"ur Physik, Humboldt Universit\"at zu Berlin, Newtonstr. 15, 12489 Berlin, Germany

      $^{(\gre)}$ Theory Group, Lab. de Physique Subatomique et de Cosmologie, 38026 Grenoble, France

      $^{(\bon)}$ HISKP (Theory), Rheinische Friedrich-Wilhelms Universit\"at Bonn, Germany

      $^{(\ber)}$ Albert Einstein Center for Fund. Physics, University of Bern, CH-3012 Bern, Switzerland

} 

\abstract{In this contribution, results from $N_f=2$ lattice QCD simulations at one lattice spacing using twisted mass fermions with a clover term at the physical pion mass are presented. The mass splitting between charged and neutral pions (including the disconnected contribution) is shown to be around $20(20)~\mathrm{MeV}$.
Further, a first measurement using the clover twisted mass action of the average momentum fraction of the pion is given. Finally, an analysis of pseudoscalar meson masses and decay constants is presented involving linear interpolations in strange and charm quark masses. Matching to meson mass ratios allows the calculation of quark mass ratios: $\mu_s/\mu_l=27.63(13)$, $\mu_c/\mu_l=339.6(2.2)$ and $\mu_c/\mu_s=12.29(10)$.
From this mass matching the quantities $f_K=153.9(7.5)~\mathrm{MeV}$, $f_D=219(11)~\mathrm{MeV}$, $f_{D_s}=255(12)~\mathrm{MeV}$ and $M_{D_s}=1894(93)~\mathrm{MeV}$ are determined without the application of finite volume or discretization artefact corrections and with errors dominated by a preliminary estimate of the lattice spacing.
}

\FullConference{The 32nd International Symposium on Lattice Field Theory,\\
		23-28 June, 2014\\
		Columbia University New York, NY}

\begin{document}

\section{Introduction}

First results from simulations employing $N_f=2$ twisted mass fermions~\cite{Frezzotti:1999vv, Frezzotti:2003ni} at the physical pion mass including a clover term have been presented in ref.~\cite{Abdel-Rehim:2013yaa} with a focus on demonstrating their feasibility.
The reader is referred there for details of the action and the simulation parameters.
In the present contribution, these results are extended by a determination of isospin symmetry violating lattice artefacts in the pion and baryon sectors.
In both cases a significant reduction is seen in comparison to $N_f=2+1+1$ simulations with Iwasaki gauge action without a clover term.
Further, a first computation of the average quark momentum fraction of the pion using the twisted mass clover action is given.
Finally, an analysis of pseudoscalar meson masses and decay constants is presented with extended statistics and an attempt is made to increase precision via the matching of meson mass ratios to their phenomenologically determined values.

\section{Estimate of the Lattice Spacing}\label{sec:lattice_spacing}

An estimate of the lattice spacing for the ensembles used in this contribution ($\beta=2.1$) can be computed  at the physical pion mass from the pion decay constant, the Sommer scale $r_0$~\cite{Sommer:1993ce} and the length scales $\sqrt{t_0}$~\cite{Luscher:2010iy} and $w_0$~\cite{Borsanyi:2012zs} calculated in the gradient flow framework with a symmetric definition of the energy density.
For completeness, determinations of these quantities are given also for non-physical pion masses in table \ref{tab:lattice_spacing}.
Taking the four values at $a\mu_l=0.0009$, their mean:
\begin{equation}
 a = 0.0937(46)~\mathrm{fm,}
 \label{eq:lattice_spacing}
\end{equation}
will serve as a rough determination of the lattice spacing in what follows.
The standard deviation is given as a generous error estimate to account for the spread of the four estimates.

\begin{table}[b]
 \hspace{-0.2cm}
 \footnotesize \centering
 \begin{tabular}{rcccc}
  \hline \hline
   & $af_\pi$ & $r_0/a$ & $\sqrt{t_0}/a$ & $w_0/a$ \\
  \hline
  $a\mu_l=0.006$ & $0.0698(3)$ & $5.162(53)$ & $1.6554(22)$ & $1.8142(41)$ \\
  $a\mu_l=0.003$ & $0.0624(5)$ & $5.322(114)$ & $1.6540(25)$ & $1.8157(55)$ \\
  $a\mu_l=0.0009$ & $0.0605(2)$ & $5.317(48)$ & $1.67443(70)$ & $1.8572(14)$ \\
  \hline
  $a~[\mathrm{fm}]$ & $0.0916(4)$ & $0.089(3)$ & $0.0997(30)$ & $0.0945(10)$ \\
  \hline \hline
 \end{tabular}
 \caption{Determinations of the length scales $r_0/a$, $w_0/a$ and $\sqrt{t_0}/a$ for three valence quark masses. The lattice spacing is computed at $a\mu=0.0009$ from the pion decay constant $f_\pi=130.4(2)~\mathrm{MeV}$ and the three length scales by setting $r_0=0.474(14)~\mathrm{fm}$~\cite{Carrasco:2014cwa}, $\sqrt{t_0}=0.167(5)~\mathrm{fm}$~\cite{Luscher:2010iy} and $w_0=0.1755(19)~\mathrm{fm}$~\cite{Borsanyi:2012zs}. Note that the scale determinations for the ensemble at $a\mu_l=0.003$ were not very stable and have comparatively large errors.}
 \label{tab:lattice_spacing}
\end{table}

\section{Isospin Symmetry Violation}

Twisted mass lattice QCD (tmLQCD) offers the advantage of automatic $\mathcal{O}(a)$ improvement at maximal twist.
However, it breaks flavour symmetry explicitly through an $\mathcal{O}(a^2)$ lattice artefact.
Although for most quantities this breaking is undetectable within errors, it is sizeable for the neutral pion, rendering it lighter than the charged ones at non-zero lattice spacing~\cite{Michael:2007vn,Dimopoulos:2009qv}.
It has been shown theoretically and practically in refs.~\cite{Colangelo:2010cu,Bar:2010jk,Carrasco:2014cwa}  that this effect should be taken into account in Wilson $\chi\mathrm{PT}$ analyses of lattice data through the introduction of appropriate $\mathcal{O}(a^2)$ terms.

\begin{figure}
 \centering
 \includegraphics[width=0.45\linewidth]{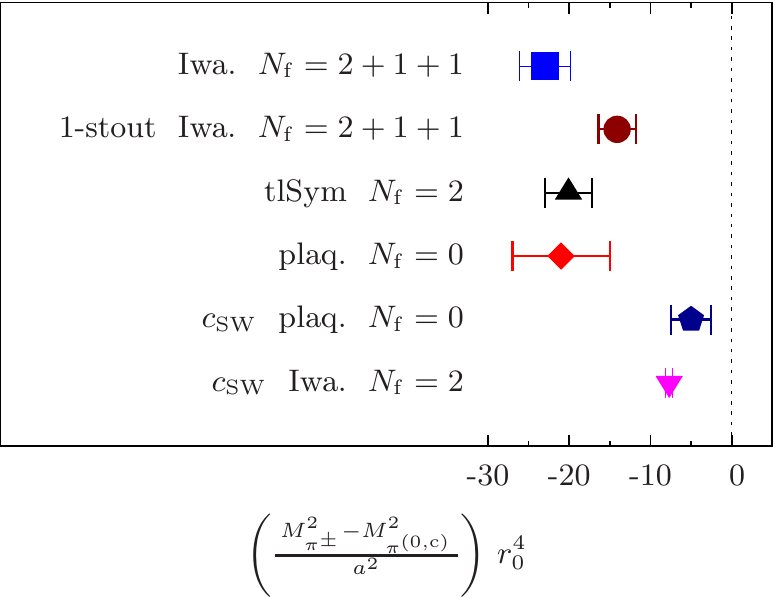}
 \includegraphics[width=0.45\linewidth]{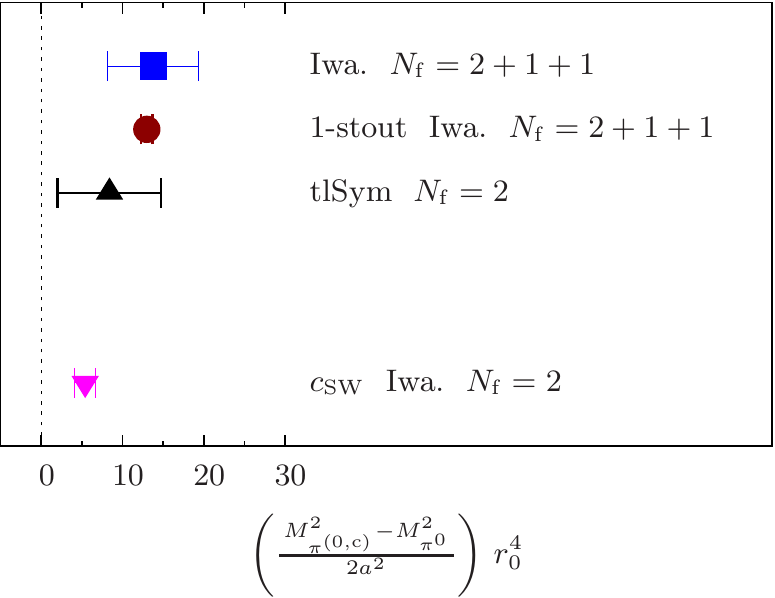}
 \includegraphics[width=0.49\linewidth]{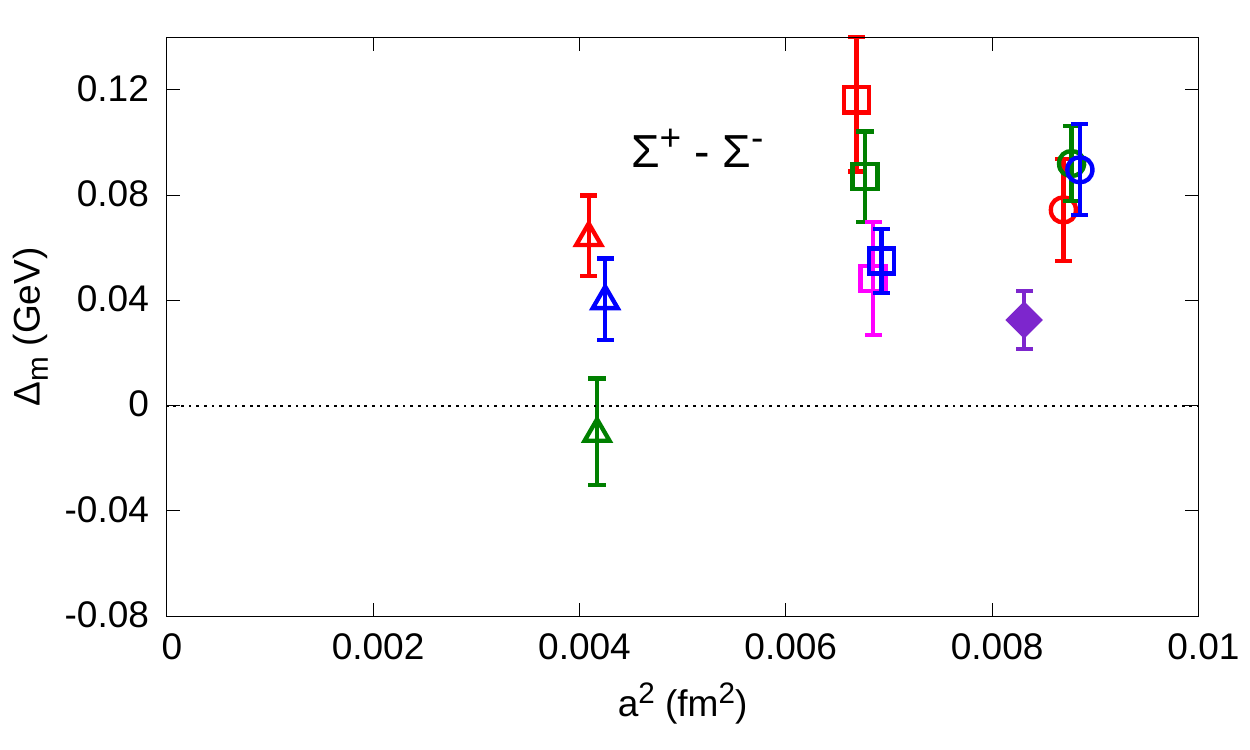}
 \includegraphics[width=0.49\linewidth]{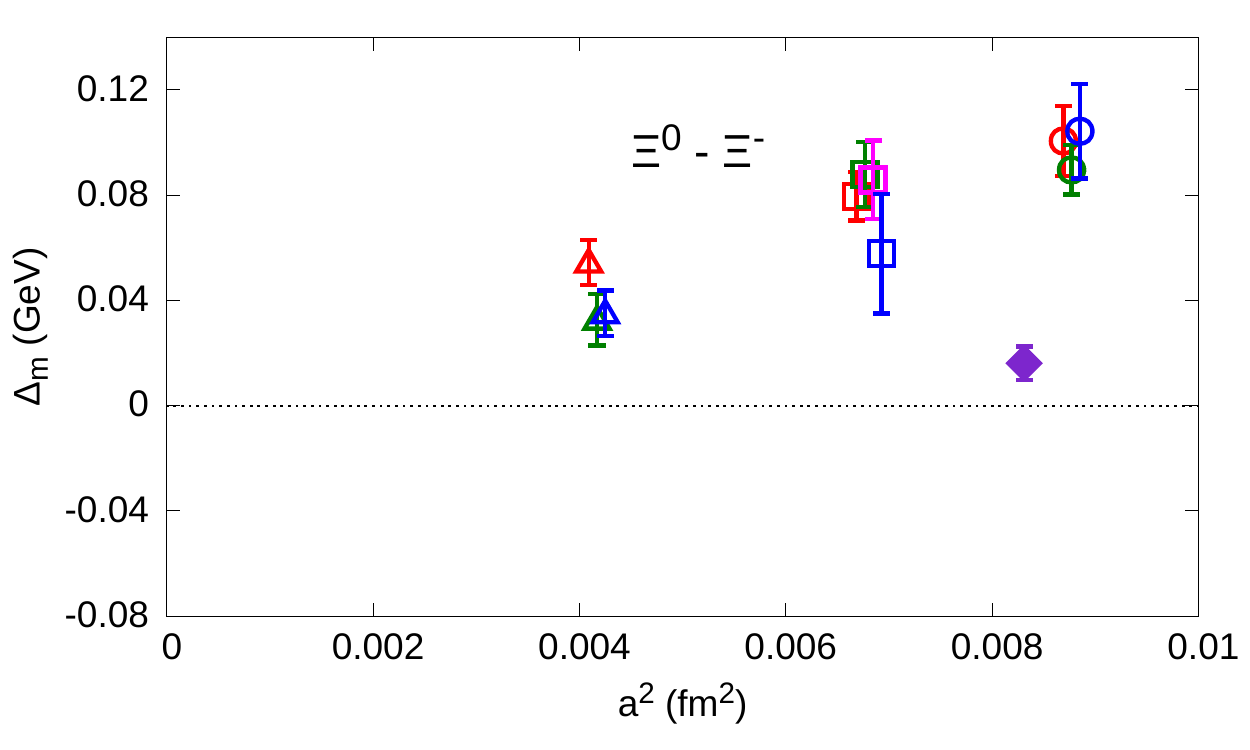}
 \caption{\protect{\bf (Top)} Difference between the charged and connected neutral pion masses (left) and the full neutral and connected neutral pion masses (right) in terms of $r_0$. The bottom-most points correspond to the present simulations. \protect{\bf(Bottom)} Mass splittings between isospin partners of $\Sigma$ and $\Xi$ baryons compared to $N_f=2+1+1$ data from ref.~\cite{Alexandrou:2014sha}. Colours correspond to different light quark masses and the filled symbol corresponds to the physical point ensemble.}
 \label{fig:mass_splitting}
\end{figure}

The mass splitting involving the full neutral pion correlator and the connected part only is shown at the top of figure \ref{fig:mass_splitting} in units of $r_0$, allowing comparison to data presented in ref.~\cite{Herdoiza:2013sla}.
This measurement was carried out on a $24^3\cdot48$ lattice with a charged pion mass of around $350~\mathrm{MeV}$ as the absolute value of the splitting depends only mildly on the quark mass.
In physical units, using the estimate of the lattice spacing given in section \ref{sec:lattice_spacing}, the mass difference between the charged pion and the full neutral pion is around $20(20)~\mathrm{MeV}$, where the error is dominated by significant stochastic noise coming from the disconnected contribution to $\pi^0$.
This corresponds to a reduction by about a factor of 5 when compared to the simulations on the coarsest lattices employed by the ETMC.
Two other quantities affected by isospin breaking are the masses of the $\Sigma$ and $\Xi$ baryon isospin partners, a preliminary determination of which indicates that the effect seems to be mildened by the twisted mass clover action, as shown at the bottom of figure \ref{fig:mass_splitting} compared to data from~\cite{Alexandrou:2014sha}.
This confirms the indication from the stability of the simulations and the connected neutral pion mass splitting presented in ref.~\cite{Abdel-Rehim:2013yaa}.

In past simulations, lattice artefacts relating the Sharpe-Singleton first-order scenario~\cite{Sharpe:1998xm} and the pion mass splitting effectively limited twisted mass fermions to light quark masses in excess of $\mathcal{O}(a^2 \Lambda_\mathrm{QCD}^3)$, corresponding to charged pions with masses of around $210~\mathrm{MeV}$ for the finest lattice spacing employed by the ETMC.
It is therefore particularly important that this splitting seems to be under much better control with the clover twisted mass action.
In some sense, this was already understood in ref.~\cite{Frezzotti:2005gi}, which proposed the addition of a non-perturbatively tuned clover term as an alternative to the ``optimally'' tuned critical mass parameter to cure the appearance of certain lattice artefacts in the pion decay constant.
This was studied numerically in the quenched approximation in ref.~\cite{Becirevic:2006ii}, in which it became clear that the connected contribution to the pion mass splitting was reduced by the addition of the clover term.
From present simulations it appears that a substantial reduction in discretization effects is achieved through the combination of a clover term (not necessarily non-perturbatively tuned) and the optimal choice of the critical mass, at least for quantities considered here at the present lattice spacing and pion mass.
Future computations need to be performed to test this for other quantities.

\section{Quark Momentum Fraction of the Pion}

\begin{wrapfigure}{r}{0.40\linewidth}
 \includegraphics[width=\linewidth]{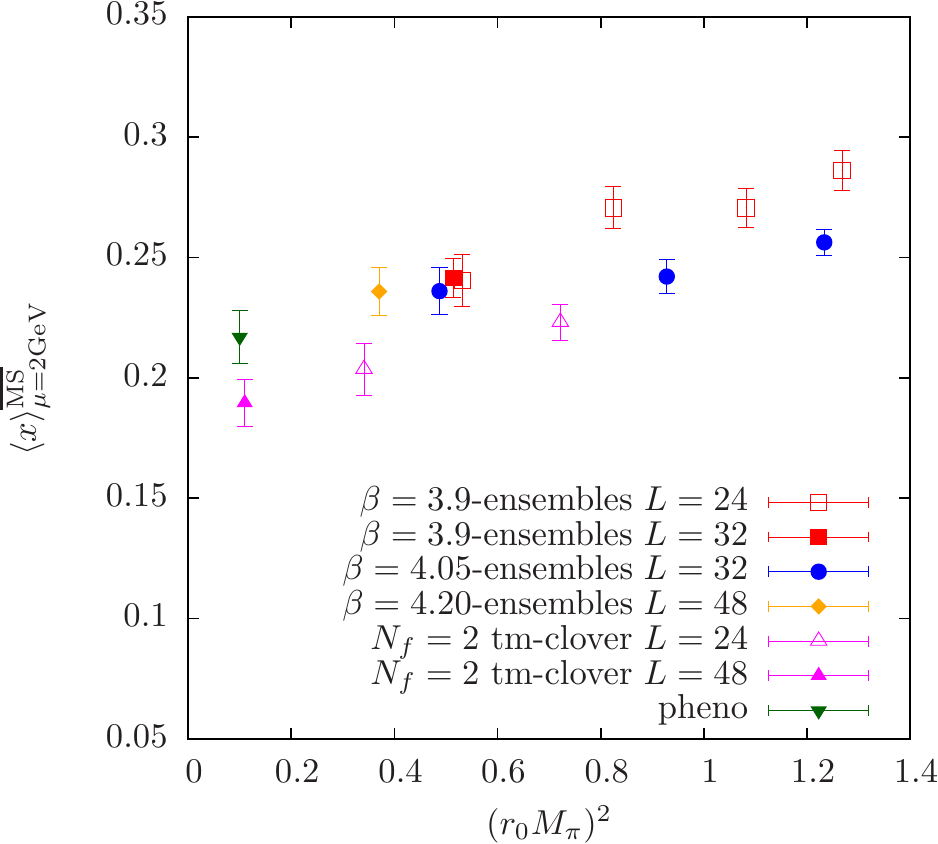}
 \caption{Average quark momentum fraction of the pion.}
 \label{fig:pion_averx}
\end{wrapfigure}

For a number of quantities there exist some disagreements between lattice results extrapolated to the physical pion mass and phenomenological determinations, indicating either some systematic effect which is not being taken into account or a discrepancy between theory and phenomenology.
One of these quantities is the average quark momentum fraction of the pion, denoted here as $\langle x \rangle$.
A measurement at the physical point, renormalized at a scale of $2~\mathrm{GeV}$ in the $\overline{\mathrm{MS}}$ scheme is shown in figure \ref{fig:pion_averx} compared to old data from ref.~\cite{Baron:2007ti} ($\beta=3.9$) and a phenomenological value from ref.~\cite{Wijesooriya:2005ir}.
This is a first step towards a full study including continuum and infinite volume limits which should allow for any remaining systematic effects to be isolated.

\section{Pseudoscalar Meson Masses and Decay Constants}\label{sec:pseudoscalar}

In twisted mass lattice QCD at maximal twist, pseudoscalar meson decay constants can be computed without further renormalization constants from the relation
\begin{equation}
 f_{\mathrm{PS}} = \frac{\left( \mu_1 + \mu_2 \right)}{M_{\mathrm{PS}}^2} \langle 0 \left| \mathcal{P} \right| \mathrm{PS} \rangle + \mathcal{O}(a^2),
 \label{eq:fps}
\end{equation}
in which the lattice dispersion relation for mesons can be taken into account by exchanging $M_\mathrm{PS}^2$ in equation \eqref{eq:fps} for $M_\mathrm{PS} \sinh(M_\mathrm{PS})$.
In the following, the former will be referred to as ``Continuum Definition'' (CD) and the latter as ``Lattice Definition'' (LD).

In ref.~\cite{Abdel-Rehim:2013yaa}, a first analysis of pseudoscalar meson masses and decay constants was presented with strange and charm quark masses tuned according to the $N_f=2$ strange to light quark mass ratio from ref.~\cite{2013:FlagReview} and the charm to strange quark mass ratio from ref.~\cite{Davies:2009ih}.
Due to limited statistics and incomplete error analysis, no explicit values were given.
For the present contribution, the analysis was extended to over 1000 measurements and strange and charm propagators were computed at four values of their respective bare mass parameters.
A full listing is given in table \ref{tab:ps_params} including fit ranges which were chosen to start where the effective masses from local-local, local-fuzzed and fuzzed-fuzzed correlators converge.

For each ensemble, the quark propagators for all valence quark masses were computed from the same stochastic local and fuzzed timeslice sources with spin dilution.
A functional form $\frac{1}{2m} A_{lf} \lbrace e^{-mt} + e^{-m(T-t)} \rbrace$ was fitted to the correlators, taking into account correlations between timeslices as well as fuzzed and local amplitudes via the inverse of the variance-covariance matrix computed according to ref.~\cite{Michael:1994sz}.
The square root of the local-local amplitude, $\sqrt{A_{ll}}$, corresponds to the matrix element from equation \eqref{eq:fps}.
The quantities given in tables \ref{tab:ps_ratios} and \ref{tab:ps_values} were analysed in a stationary bootstrap~\cite{stationarybootstrap} framework with a mean block length chosen to accommodate autocorrelations between measurements.

\begin{table}
 \caption{Fit ranges and bare strange and charm quark masses for the computation of pseudoscalar meson correlators.}
 \label{tab:ps_params}
 {\footnotesize
 \begin{tabular}{|l|ll|ccccc|}
  \hline
  \multirow{2}{*}{$L/a$} & \multicolumn{2}{|c|}{\multirow{2}{*}{bare valence quark masses}} & \multicolumn{5}{c|}{fit ranges} \\
  & & & $\pi^\pm$ & $\pi^0$ & $K$ & $D$ & $D_s$ \\
  \hline
  \multirow{3}{*}{24} & $a\mu_l$ & $0.003, 0.006$ & \multirow{3}{*}{$[10,23]$} & \multirow{3}{*}{$[11,20]$} & \multirow{3}{*}{$[12,20]$} & \multirow{3}{*}{$[15,20]$} & \multirow{3}{*}{$[16,20]$} \\
  & $a\mu_s$ & $0.0224, 0.0231, 0.0238, 0.0245, 0.0252, 0.0259$ & & & & & \\
  & $a\mu_c$ & $0.2586, 0.2704, 0.2822, 0.294, 0.3058, 0.3176$ & & & & & \\
  \hline
  \multirow{3}{*}{48} & $a\mu_l$ & $0.0009$ & \multirow{3}{*}{$[10,42]$} & \multirow{3}{*}{$[11,36]$} & \multirow{3}{*}{$[12,42]$} & \multirow{3}{*}{$[15,30]$} & \multirow{3}{*}{$[18,32]$} \\
  & $a\mu_s$ & $0.0231, 0.0238, 0.0245, 0.0252$ & & & & & \\
  & $a\mu_c$ & $0.2704, 0.2822, 0.294, 0.3058$ & & & & & \\
  \hline
 \end{tabular}
 \vspace{0.2cm}
 } 
\end{table}

At $a\mu_l=0.0009$, corresponding to the physical pion mass, the quantities were fitted to simple linear models of the form $\alpha(a\mu_s)+\beta(a\mu_c)+\gamma$ to interpolate to physical strange and charm quark masses as determined from the matching procedure described below.
The quoted errors have a statistical contribution from the bootstrap procedure and an error coming from the fit added in quadrature.
The latter includes the propagation via first-order Taylor expansion of the error on the estimate of the physical strange and charm quark masses, which can be quite sizeable for meson masses (as can be seen in figure \ref{fig:mass_matching}) but is negligible for decay constants.

\subsection{Strange and Charm Mass Matching}\label{sec:mass_matching}

\begin{figure}
 \centering
 \includegraphics[width=0.46\linewidth]{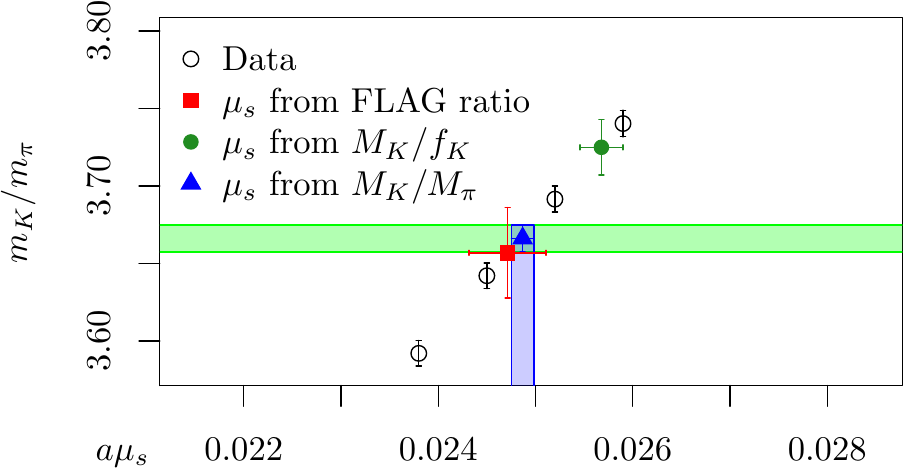}
 \includegraphics[width=0.46\linewidth]{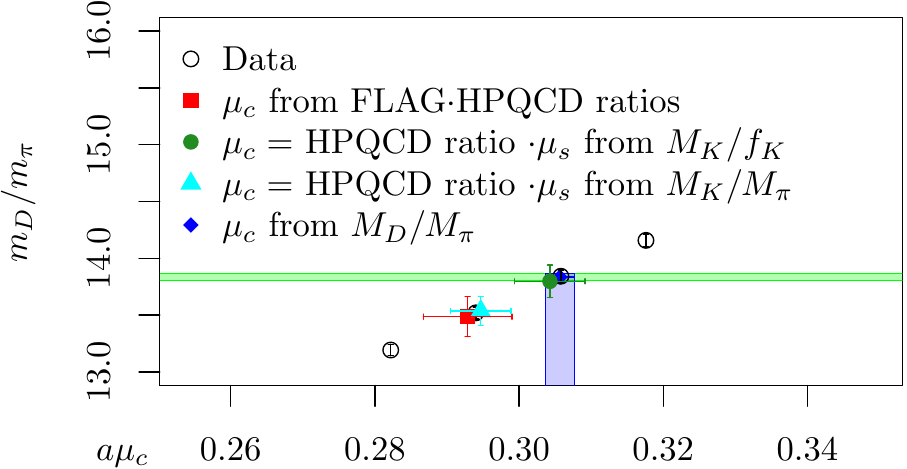}
 \caption{Linear interpolations in the valence strange and charm masses of the ratios $M_K/M_\pi$ and $M_D/M_\pi$. The green band indicates the phenomenological value and the blue band the result of the matching. The other points are for reference and show the effect of using different matching conditions or relying on quark mass ratios as input.}
 \label{fig:mass_matching}
\end{figure}

In order to obtain precise estimates of the bare strange and charm quark masses corresponding to physical renormalized quark masses, the ratios $M_K/M_\pi$ and $M_D/M_\pi$ were interpolated as described above and matched to the isospin symmetric value of $M_K/M_\pi$ from ref.~\cite{2013:FlagReview} and $M_D$ from ref.~\cite{2012:PDG} with $M_\pi$ still in the isospin symmetric limit.
The results of this matching procedure, illustrated in figure \ref{fig:mass_matching}, are given in bold in table \ref{tab:mass_matching} in addition to $a\mu_s$ derived from matching $M_K/f_K$ as well as $a\mu_s$ from the $N_f=2+1$ strange to light quark mass ratio of ref.~\cite{2013:FlagReview}.
The quark mass ratios corresponding to the $M_K$ and $M_D$ matching are given in table \ref{tab:mass_ratios}.

\begin{table}
 \begin{subtable}{0.6\linewidth}
  \subcaption{Bare quark masses resulting from matching using the quantity in the table header. The labels (LD) and (CD) correspond to $f_K$ extracted according to the two definitions given in equation \eqref{eq:fps}. The starred $a\mu_c$ are derived from the corresponding $a\mu_s$ and the HPQCD c to s ratio.}
  \label{tab:mass_matching}
  \footnotesize
  \begin{tabular}{cccccc}
    \hline \hline
    & FLAG & $M_K/f_K^\mathrm{(CD)}$ & $M_K/f_K^\mathrm{(LD)}$ & $M_K/M_\pi$ & $M_D/M_\pi$ \\
    \hline
    $a\mu_s$ & $0.0247(4)$ & $0.0257(2)$ & $0.0251(2)$ & $\mathbf{0.0249(1)}$ & -- \\
    $a\mu_c$ & $0.293(6)^\star$ & $0.305(5)^\star$ & $0.0297(5)^\star$ & $0.295(4)^\star$ & $\mathbf{0.306(2)}$ \\
    \hline \hline
  \end{tabular}
 \end{subtable}
 \hspace{0.03\textwidth}
 \begin{subtable}{0.37\linewidth}
  \footnotesize
  \subcaption{Quark mass ratios derived from the quark masses matched through $M_K/M_\pi$ and $M_D/M_\pi$.}
  \label{tab:mass_ratios}
  \centering
  \begin{tabular}{ccc}
    \hline \hline
    $\mu_s/\mu_l$ & $\mu_c/\mu_l$ & $\mu_c/\mu_s$ \\
    \hline
    $27.63(13)$ & $339.6(2.2)$ & $12.29(10)$ \\
    \hline \hline 
  \end{tabular}
 \end{subtable} \\
 \vspace{0.2cm}
 \stepcounter{table}
\end{table}

\subsection{Results}

\begin{table}
 \footnotesize
 \hspace{-1.3cm}
 \begin{tabular}{cllllllllll}
 \hline \hline
 & $M_\pi/f_\pi$ & $M_K/f_K$ & $M_{D_s}/f_{D_s}$ & $M_{D_s}/M_\pi$ & $f_K/f_\pi$ & $f_D/f_\pi$ & $f_{D_s}/f_\pi$ & $f_D/f_K$ & $f_{D_s}/f_K$ & $f_{D_s}/f_D$ \\
 \hline
 $\mathrm{lat.^{(CD)}}$ & $1.026(5)^\dagger$ & $3.127(11)^\dagger$ & $7.44(3)$ & $14.48(6)$ & $1.203(5)$ & $1.72(3)$ & $1.996(10)$ & $1.43(2)$ & $1.658(6)$ & $1.16(2)$ \\
 $\mathrm{lat.^{(LD)}}$ & $1.027(6)^\dagger$ & $3.155(11)^\dagger$ & $8.49(4)$ & -- & $1.194(5)$ & $1.53(2)$ & $1.751(9)$ & $1.28(2)$ & $1.467(5)$ & $1.148(16)$ \\
 PDG & $1.034(3)^\star$ & $3.164(14)^\star$ & $7.64(14)$ & $14.60(3)^\star$ & $1.198(6)$ & $1.57(4)$ & $1.97(4)$ & $1.31(3)$ & $1.65(3)$ & $1.26(4)$ \\
 FLAG & $1.035(11)^\star$ & $3.162(18)^\star$ & -- & -- & $1.200(15)$ & $1.61(3)$ & $1.91(3)$ & $1.34(2)$ & $1.59(2)$ & $1.19(2)$ \\
 \hline \hline
 \end{tabular}
\caption{Ratios of pseudoscalar meson observables calculated on the gauge ensemble at the physical pion mass interpolated to the strange and charm valence quark masses from the matching in section \protect\ref{sec:mass_matching}. For comparison, values from refs.~\protect\cite{2012:PDG,2013:FlagReview} are also given. There, starred quantities involving $M_\pi$ or $M_K$ use the isospin symmetric values of these quantities. Daggered quantities are not independent and given for reference only. ``FLAG'' refers to $N_f=2+1$ averages.}
 \label{tab:ps_ratios}
\end{table}

Interpolated to the quark masses as described above, ratios of masses and decay constants are given in table \ref{tab:ps_ratios}.
It is clear from $M_\pi/f_\pi$ that the ensemble is at the physical pion mass, up to possibly sizeable finite volume effects which will be studied in the near future.
As an estimate of the residual $\mathcal{O}(a^2)$ artefacts, one can compare the difference between the two definitions of the decay constant in quantities involving $f_D$ and $f_{D_s}$.
It seems that these effects should be no larger than about 15\%, indicating that a well-behaved continuum limit should certainly be achievable.
Finally, a preliminary determination of $f_K$, $f_D$, $f_{D_s}$ and $M_{D_s}$ in physical units is given in table \ref{tab:ps_values}.
It should be noted that none of these results have been corrected for finite-size or discretization effects.

\begin{table}[b]
 \centering
 \footnotesize
 \begin{tabular}{rllll}
  \hline \hline
  & $f_K$ & $f_D$ & $f_{D_s}$ & $M_{D_s}$ \\
  \hline
  $\mathrm{lat.^{(CD)}}$ & $0.07280(17)$ & $0.1041(16)$ & $0.1208(4)$ & $0.899(3)$ \\
  $\mathrm{lat.^{(LD)}}$ & $0.07218(17)$ & $0.0923(1)$ & $0.1059(3)$ & -- \\
  $\mathrm{lat.^{(CD)}}$ (MeV) & $153.9(7.5)$ & $219(11)$ & $255(12)$ & $1894(93)$\\
  PDG (MeV) & $156.2(7)$ & $204.6(5.0)$ & $257.5(4.6)$ & $1968.50(32)$ \\
  FLAG (MeV) & $156.3(9)$ & $209.2(3.3)$ & $248.6(2.7)$ & -- \\
  \hline \hline
 \end{tabular}
 \caption{Lattice predictions calculated on the gauge ensemble at the physical pion mass, converted to physical units by means of the estimate of the lattice spacing from section \protect\ref{sec:lattice_spacing}. ``FLAG'' refers to $N_f=2+1$ averages. No finite-size or discretization artefact corrections have been applied. }
 \label{tab:ps_values}
\end{table}

\section{Conclusions and Outlook}

The results presented in this contribution confirm the suitability of the twisted mass clover action for simulations at the physical point.
It has been shown that flavour symmetry violating lattice artefacts in the pion and baryon sectors are significantly reduced compared to simulations without a clover term, supporting the first impression from the stability of the simulations.

A first computation using this action of the average momentum fraction of the pion has been presented and eventual continuum and infinite volume limits should be able to identify remaining systematic effects.
In the pseudoscalar meson sector, a mass matching procedure was explored at the physical pion mass with the result that partially quenched strange and charm mass parameters can be determined with high precision, allowing a computation of quark mass ratios largely consistent with other sources.
The values as well as ratios of pseudoscalar meson masses and decay constants were interpolated linearly to these strange and charm masses and results mostly consistent with phenomenology were found, although the currently large error on the estimate of the lattice spacing and the lack of continuum and infinite volume limits should be kept in mind.
A further source of systematic error related to the choice of fit range will be addressed in a more complete publication and the remaining discrepancies in quantities involving charm quarks will be addressed in the near future based on simulations on larger lattices and multiple lattice spacings.

{ \small
{\bf Acknowledgements: } B.K. acknowledges full support by the National Research Fund, Luxembourg under AFR Ph.D. grant 27773315.
This project was partly funded by the DFG as a project in the Sino-German CRC 110.
The calculations for this contribution were done on JuQueen at JSC in Juelich with grants provided through PRACE and the Gauss centre for supercomputing.
}

\bibliographystyle{3authors_notitle}
\bibliography{bibliography}

\begin{thebibliography}{10}
\providecommand{\url}[1]{\texttt{#1}}
\providecommand{\urlprefix}{URL }
\providecommand{\eprint}[2][]{\url{#2}}

\bibitem{Frezzotti:1999vv}
R.~Frezzotti \emph{et~al.}
\newblock Nucl.Phys.Proc.Suppl. (2000).
\newblock 83:941--946.
\newblock \eprint{hep-lat/9909003}

\bibitem{Frezzotti:2003ni}
R.~Frezzotti and G.~Rossi.
\newblock JHEP (2004).
\newblock 0408:007.
\newblock \eprint{hep-lat/0306014}

\bibitem{Abdel-Rehim:2013yaa}
A.~Abdel-Rehim \emph{et~al.}
\newblock PoS (2013).
\newblock LATTICE2013:264.
\newblock \eprint{1311.4522}

\bibitem{Sommer:1993ce}
R.~Sommer.
\newblock Nucl. Phys. (1994).
\newblock B411:839--854.
\newblock \eprint{hep-lat/9310022}

\bibitem{Luscher:2010iy}
M.~Luscher.
\newblock JHEP (2010).
\newblock 1008:071.
\newblock \eprint{1006.4518}

\bibitem{Borsanyi:2012zs}
S.~Borsanyi \emph{et~al.}
\newblock JHEP (2012).
\newblock 1209:010.
\newblock \eprint{1203.4469}

\bibitem{Carrasco:2014cwa}
N.~Carrasco \emph{et~al.} (ETM Collaboration).
\newblock Nucl.Phys. (2014).
\newblock B887:19--68.
\newblock \eprint{1403.4504}

\bibitem{Michael:2007vn}
C.~Michael and C.~Urbach (ETM Collaboration).
\newblock PoS (2007).
\newblock LAT2007:122.
\newblock \eprint{0709.4564}

\bibitem{Dimopoulos:2009qv}
P.~Dimopoulos \emph{et~al.} (ETM).
\newblock Phys.Rev. (2010).
\newblock D81:034509.
\newblock \eprint{0908.0451}

\bibitem{Colangelo:2010cu}
G.~Colangelo, U.~Wenger and J.~M. Wu.
\newblock Phys.Rev. (2010).
\newblock D82:034502.
\newblock \eprint{1003.0847}

\bibitem{Bar:2010jk}
O.~B{\"a}r.
\newblock Phys.Rev. (2010).
\newblock D82:094505.
\newblock \eprint{1008.0784}

\bibitem{Herdoiza:2013sla}
G.~Herdoiza \emph{et~al.} (ETM).
\newblock JHEP (2013).
\newblock 1305:038.
\newblock \eprint{1303.3516}

\bibitem{Alexandrou:2014sha}
C.~Alexandrou \emph{et~al.} (2014).
\newblock \eprint{1406.4310}

\bibitem{Sharpe:1998xm}
S.~R. Sharpe and R.~J. Singleton.
\newblock Phys. Rev. (1998).
\newblock D58:074501.
\newblock \eprint{hep-lat/9804028}

\bibitem{Frezzotti:2005gi}
R.~Frezzotti \emph{et~al.}
\newblock JHEP (2006).
\newblock 0604:038.
\newblock \eprint{hep-lat/0503034}

\bibitem{Becirevic:2006ii}
D.~Becirevic \emph{et~al.}
\newblock Phys.Rev. (2006).
\newblock D74:034501.
\newblock \eprint{hep-lat/0605006}

\bibitem{Baron:2007ti}
R.~Baron \emph{et~al.} (ETM Collaboration).
\newblock PoS (2007).
\newblock LAT2007:153.
\newblock \eprint{0710.1580}

\bibitem{Wijesooriya:2005ir}
K.~Wijesooriya, P.~Reimer and R.~Holt.
\newblock Phys.Rev. (2005).
\newblock C72:065203.
\newblock \eprint{nucl-ex/0509012}

\bibitem{2013:FlagReview}
S.~Aoki \emph{et~al.}
\newblock Eur.Phys.J. (2014).
\newblock C74(9):2890.
\newblock \eprint{1310.8555}

\bibitem{Davies:2009ih}
C.~Davies \emph{et~al.}
\newblock Phys.Rev.Lett. (2010).
\newblock 104:132003.
\newblock \eprint{0910.3102}

\bibitem{Michael:1994sz}
C.~Michael and A.~McKerrell.
\newblock Phys.Rev. (1995).
\newblock D51:3745--3750.
\newblock \eprint{hep-lat/9412087}

\bibitem{stationarybootstrap}
D.~N. Politis and J.~P. Romano.
\newblock J. of the American Stat. Assoc. (1994).
\newblock 89(428):1303--1313

\bibitem{2012:PDG}
J.~Beringer \emph{et~al.} (Particle Data Group).
\newblock Phys. Rev. D (2012).
\newblock 86:010001

\end{thebibliography}

\end{document}